\documentclass[12pt]{article}
\usepackage{amsmath}
\usepackage{enumerate}
\usepackage{natbib}
\usepackage{url} 
\usepackage{comment}

\newcommand{\blind}{1}

\usepackage[toc,page]{appendix}
\usepackage{amsmath}
\usepackage{multirow}

\usepackage{graphicx}

\usepackage{arydshln}
\usepackage{amssymb}
\usepackage{float}
\usepackage{subcaption}
\usepackage{rotating}  
\usepackage{csquotes}

\newcommand{\erf}{\mathrm{erf}}
\allowdisplaybreaks

\begin{document}

\def\spacingset#1{\renewcommand{\baselinestretch}%
{#1}\small\normalsize} \spacingset{1}


\if1\blind
{
  \title{\bf Zipf's law in the distribution of Brazilian firm size}

 \author{Thiago Trafane Oliveira Santos\\
    Central Bank of Brazil, Brasília, Brazil\\ Department of Economics, University of Brasilia, Brazil.\\ Email: thiago.trafane@bcb.gov.br\\
    and \\
    Daniel Oliveira Cajueiro \\
    Department of Economics, University of Brasilia, Brazil.\\  National Institute of Science and Technology for Complex Systems (INCT-SC)\\Machine Learning Laboratory in Finance and Organizations (LAMFO), Brazil.\\ Email: danielcajueiro@gmail.com}
  \maketitle
} \fi

\if1\blind
{
  \bigskip
  \bigskip
  \bigskip
  \begin{center}
\end{center}
  \medskip
} \fi

\bigskip
\begin{abstract}
			Zipf's law states that the probability of a variable being larger than $s$ is roughly inversely proportional to $s$. In this paper, we evaluate Zipf's law for the distribution of firm size by the number of employees in Brazil. We use publicly available binned annual data from the Central Register of Enterprises (CEMPRE), which is held by the Brazilian Institute of Geography and Statistics (IBGE) and covers all formal organizations. Remarkably, we find that Zipf's law provides a very good, although not perfect, approximation to data for each year between 1996 and 2020 at the economy-wide level and also for agriculture, industry, and services alone. However, a lognormal distribution also performs well and even outperforms Zipf's law in certain cases.\end{abstract}

\noindent%
\vfill

\newpage
\spacingset{1.8} 

	\section{Introduction} \label{sec:intro}
	
	A power or scaling law holds for variables $X$ and $Y$ if $Y = c X^k$, where $k$ is known as the power law exponent and $c$ is typically an unremarkable constant. As \citet{gabaix2009power,gabaix2016power} points out, these power laws emerge in different domains, from natural phenomena (e.g., earthquakes, forest fires, and rivers), biology (e.g., Kleiber’s law), and popularity of websites to economics, both in theory (e.g., the quantity theory of money) and empirically (e.g., Kaldor’s stylized facts on economic growth). A power law may also apply to a distribution, with 
	\begin{align} 
		P(S\geq s) =  \left(\underline{s}/s\right)^k \label{eq:sf_pareto}
	\end{align}	
	for a random variable $S$, $S\geq\underline{s}>0$, where $k>0$. Generally, this distribution is known as Pareto (type I), but it is called Zipf's law when $k\approx1$. In such cases, the probability of $S$ being greater or equal to $s$ is roughly proportional to $1/s$. This ``law'' was named after the linguist George Kingsley Zipf, who found analogous empirical regularity for the usage frequency of words in different languages and countries \citep{zipf1949human}, but it shows up in several other contexts. One illustrative example is the distribution of city size by population, especially among larger cities \citep{gabaix1999zipf,gabaix_ioannides2004evolution}.\footnote{We may find analogous evidence for Brazil \citep{mourajr_ribeiro2006zipf,justo2014zipf}. On a related matter, \citet{comitti_etal2022days} estimate daily power law exponent $k$ for the distribution of Brazilian municipalities by the number of infected people by COVID-19. Interestingly, they find it converges over time to $0.87$, which is exactly the $k$ they estimate for the distribution of municipality size by population.}
	
	In this paper, we evaluate Zipf's law for the distribution of firm size by the number of employees in Brazil. We use publicly available binned annual data from the Central Register of Enterprises (CEMPRE), which is held by the Brazilian Institute of Geography and Statistics (IBGE) and covers all formal organizations. Following the methodology proposed by  \citet{virkar_clauset2014power}, we find that Zipf's law provides a very good, although not perfect, approximation to data for each year between 1996 and 2020 at the economy-wide level and also for agriculture, industry, and services alone. However, a lognormal distribution also performs well and even outperforms Zipf's law in certain cases.
	
	Empirical evidence supporting Zipf's law for firm size distribution has been found for several different countries with firm size measured by the number of employees, sales, income, total assets, and equity plus debt \citep{okuyama_etal1999zipf,axtell2001zipf,fujiwara_etal2004pareto,luttmer2007selection,gabaix_landier2008why,digiovanni_etal2011power,digiovanni_levchenko2013firm}.\footnote{\citet{fujiwara2004zipf} finds that Zipf's law also holds for the distribution of total liabilities of \textit{bankrupted} firms in Japan. For a survey of the empirical findings about Zipf's law for firm size, see Section 3 of  \citet{bottazzi_etal2015zipf}.} In particular,  \citet{digiovanni_levchenko2013firm} use the ORBIS database to evaluate firm size distribution by total sales for a sample of 44 countries. In their own words, \blockquote{[...] the country sample is diverse: it includes major European economies (France, Germany, Netherlands), smaller E.U. accession countries (Czech Republic, Estonia), major middle income countries (Brazil, Argentina), as well as the two largest emerging markets (India and China). All in all, in this sample of 44 countries with very different characteristics, the distributions of firm size are remarkably consistent with Zipf's Law.} Specifically to Brazil,  \citet{dasilva_etal2018granularity} study the distribution of firm size by net revenue, finding support for Zipf's law among the 1,000 largest firms in 2015.	 
	
	The literature also shows contradictory evidence. For instance, there is some support for lognormality for firm size distribution \citep{stanley_etal1995zipf,kondo_etal2023heavy}. Moreover, applying Lagrange multiplier tests,  \citet{resende_cardoso2022firm} find support to the more general Pareto type II and Pareto type IV against the Pareto type I and Zipf's law for firm size distribution by net revenue in Brazil.
	
	\bigskip
	The remainder of the paper proceeds as follows. Section \ref{sec:data_method} presents the data and methodology. Section \ref{sec:results} presents the empirical results. Finally, Section \ref{sec:conclusion} concludes.
	
	\section{Data and methodology} \label{sec:data_method}
	
	We use publicly available annual data from CEMPRE, which is held by the IBGE and covers all formal organizations (corporate entities, public administration, and non-profit organizations). We split the analysis into two distinct periods due to a methodological break in the database, which (i) altered the criteria for identifying active firms and (ii) updated the industry classification. First, between 1996 and 2006, when the industries are classified according to the National Classification of Economic Activities (CNAE), a Brazilian classification derived from the ISIC Rev.3. Second, from 2006 to 2020, using CNAE 2.0, which follows the ISIC Rev.4. For both periods, we have the number of firms across all industries (up to 3-digit level) by nine size bins based on the number of employees: 0 to 4, 5 to 9, 10 to 19, 20 to 29, 30 to 49, 50 to 99, 100 to 249, 250 to 499, and 500 or more. All our analyses are done at the economy-wide level and also for agriculture, industry, and services alone.\footnote{For the CNAE, we classified sections A and B as agriculture, C to F as industry, and G to Q as services. For the CNAE 2.0, A is agriculture, B to F is industry, and G to U is services.} Table \ref{tab:descrip} presents the data for these industries in selected years, showing the new criteria for identifying active firms substantially lowered the number of firms in 2006, notably for firms with up to four employees.
	
	\begin{table}[p]
		\centering
		\caption{Number of firms by firm size}
		\label{tab:descrip}
        \resizebox{\textwidth}{!}{%
		\begin{tabular}{llcccccccc}
			\hline \hline
			\multicolumn{2}{l}{Number of} & &  \multicolumn{3}{c}{1996-2006 database} & & \multicolumn{3}{c}{2006-2020 database} \\ \cline{4-6} \cline{8-10} 
			\multicolumn{2}{l}{employees} & & 1996 & 2001 & 2006 & & 2006 & 2013 & 2020 \\
			\hline	
			\multicolumn{2}{l}{\textit{All industries}} \\	
			& 0 to 4 &                         &         2,616,788 &         3,903,486 &         4,730,580 &                         &         3,324,519 &         3,985,367 &         4,090,186 \\
			& 5 to 9 &                         &           327,372 &           432,626 &           542,426 &                         &           531,612 &           755,609 &           739,242 \\
			& 10 to 19 &                         &           141,337 &           193,133 &           265,581 &                         &           261,271 &           379,902 &           358,736 \\
			& 20 to 29 &                         &            40,693 &            55,032 &            69,486 &                         &            69,433 &           102,152 &            93,372 \\
			& 30 to 49 &                         &            31,260 &            39,498 &            50,276 &                         &            50,222 &            73,368 &            65,053 \\
			& 50 to 99 &                         &            23,133 &            27,102 &            33,294 &                         &            33,269 &            47,651 &            43,294 \\
			& 100 to 249 &                         &            15,244 &            16,732 &            19,683 &                         &            19,664 &            27,132 &            24,341 \\
			& 250 to 499 &                         &             5,713 &             6,283 &             7,807 &                         &             7,801 &            10,429 &             9,739 \\
			& 500 or more &                         &             5,181 &             5,933 &             7,793 &                         &             7,787 &            10,624 &            10,128 \\
			& Total &                         &         3,206,721 &         4,679,825 &         5,726,926 &                         &         4,305,578 &         5,392,234 &         5,434,091 \\
			\hline
			\multicolumn{2}{l}{\textit{Agriculture}}\textit{} \\	
			& 0 to 4 &                         &            16,419 &            23,666 &            38,961 &                         &            21,850 &            93,237 &            89,402 \\
			& 5 to 9 &                         &             3,436 &             3,737 &             4,681 &                         &             4,249 &             5,870 &             6,638 \\
			& 10 to 19 &                         &             1,909 &             2,160 &             2,948 &                         &             2,740 &             3,686 &             3,657 \\
			& 20 to 29 &                         &               735 &               814 &               980 &                         &               977 &             1,105 &             1,136 \\
			& 30 to 49 &                         &               583 &               717 &               778 &                         &               760 &               863 &               811 \\
			& 50 to 99 &                         &               447 &               538 &               585 &                         &               599 &               637 &               681 \\
			& 100 to 249 &                         &               247 &               310 &               404 &                         &               402 &               398 &               407 \\
			& 250 to 499 &                         &               103 &               132 &               121 &                         &               125 &               157 &               160 \\
			& 500 or more &                         &                88 &               124 &               127 &                         &               127 &               127 &               130 \\
			& Total &                         &            23,967 &            32,198 &            49,585 &                         &            31,829 &           106,080 &           103,022 \\
			\hline
			\multicolumn{2}{l}{\textit{Industry}}\textit{} \\	
			& 0 to 4 &                         &           315,907 &           413,192 &           474,964 &                         &           314,128 &           433,166 &           472,907 \\
			& 5 to 9 &                         &            61,262 &            73,224 &            83,092 &                         &            82,158 &           118,577 &           107,997 \\
			& 10 to 19 &                         &            36,803 &            48,727 &            59,429 &                         &            59,166 &            79,931 &            69,686 \\
			& 20 to 29 &                         &            13,656 &            18,474 &            21,407 &                         &            21,664 &            29,726 &            24,498 \\
			& 30 to 49 &                         &            11,487 &            14,795 &            17,571 &                         &            17,588 &            23,142 &            17,941 \\
			& 50 to 99 &                         &             9,045 &            10,906 &            13,200 &                         &            13,231 &            17,366 &            13,127 \\
			& 100 to 249 &                         &             5,759 &             6,160 &             7,308 &                         &             7,295 &             9,836 &             7,399 \\
			& 250 to 499 &                         &             2,089 &             1,942 &             2,438 &                         &             2,423 &             3,228 &             2,596 \\
			& 500 or more &                         &             1,726 &             1,622 &             2,034 &                         &             2,038 &             2,902 &             2,283 \\
			& Total &                         &           457,734 &           589,042 &           681,443 &                         &           519,691 &           717,874 &           718,434 \\ 
			\hline
			\multicolumn{2}{l}{\textit{Services}}\textit{} \\	
			& 0 to 4 &                         &         2,284,462 &         3,466,628 &         4,216,655 &                         &         2,988,541 &         3,458,964 &         3,527,877 \\
			& 5 to 9 &                         &           262,674 &           355,665 &           454,653 &                         &           445,205 &           631,162 &           624,607 \\
			& 10 to 19 &                         &           102,625 &           142,246 &           203,204 &                         &           199,365 &           296,285 &           285,393 \\
			& 20 to 29 &                         &            26,302 &            35,744 &            47,099 &                         &            46,792 &            71,321 &            67,738 \\
			& 30 to 49 &                         &            19,190 &            23,986 &            31,927 &                         &            31,874 &            49,363 &            46,301 \\
			& 50 to 99 &                         &            13,641 &            15,658 &            19,509 &                         &            19,439 &            29,648 &            29,486 \\
			& 100 to 249 &                         &             9,238 &            10,262 &            11,971 &                         &            11,967 &            16,898 &            16,535 \\
			& 250 to 499 &                         &             3,521 &             4,209 &             5,248 &                         &             5,253 &             7,044 &             6,983 \\
			& 500 or more &                         &             3,367 &             4,187 &             5,632 &                         &             5,622 &             7,595 &             7,715 \\
			& Total &                         &         2,725,020 &         4,058,585 &         4,995,898 &                         &         3,754,058 &         4,568,280 &         4,612,635 \\
			\hline \hline
			\multicolumn{9}{l}{Source: publicly available CEMPRE database.}
		\end{tabular}}
	\end{table}		
	
	 \citet{virkar_clauset2014power} suggest three steps to evaluate the prevalence of a distributional power law in binned data: (i) fit the power law, (ii) test the power law's plausibility, and (iii) compare against alternative distributions.\footnote{For an analogous approach for non-binned data, see  \citet{clauset_etal2009power}.} We follow similar steps. Our alternative distributions are (i) a strong Zipf's law or simply a Zipf distribution, that is, a Pareto density with $k=1$, and (ii) a lognormal density. The choice of the lognormal is due to two reasons. First, the ``[..] lognormal provides a strong test because for a wide range of sample sizes it produces bin counts that are reasonably power-law-like when plotted on log–log axes [...]'' \citep{virkar_clauset2014power}. Second, there is also evidence supporting lognormality for firm size distribution \citep{stanley_etal1995zipf,kondo_etal2023heavy}. Given these alternative distributions, we consider the following three steps to evaluate power and strong Zipf's law:	
	\begin{enumerate}
		\item Fit Pareto and lognormal distributions.
		\item Test Pareto, lognormal, and strong Zipf's law plausibility.
		\item Compare Pareto, Zipf, and lognormal distributions.
	\end{enumerate}
	In the following, we present the methodology used in each of these three steps.
	
	\subsection{Step 1: fitting the distributions} \label{sec:method_step1}
	
	Before discussing the estimators, three comments are in order. First, in some empirical applications, a distributional power law may hold but only in the upper tail, meaning one must also gauge the support's lower bound $\underline{s}>0$. For instance, one can visually identify the point beyond which the empirical survival function becomes roughly straight on a log-log plot, although more objective methods also exist \citep{breiman_etal1990robust,dekkers_dehaan1993optimal,drees_kaufmann1998selecting,danielsson_etal2001using,handcock_jones2004likelihood,clauset_etal2007frequency}. However, since we have just a few bins, we choose to test all possible $\underline{s}$ instead of choosing a specific one, setting $\underline{s} = 5,10,20,30,50$ for both Pareto and lognormal distributions.\footnote{We do not set $\underline{s}=0$ because the Pareto support is strictly positive, while $\underline{s} = 100,250,500$ are discarded as we need at least four bins to ensure some degree of freedom in the lognormal estimation.} Second, since lognormal's support begins at zero and we need it to start at $\underline{s}>0$, we shift its density to the right by $\underline{s}$, supposing $S-\underline{s}>0$ is lognormally distributed. Third, our measure of firm size, the number of employees, is discrete, whereas both Pareto and lognormal are continuous distributions. We address this issue by discretizing each distribution, defining the probability mass function as $P(S=s) \equiv P(S\geq s) - P(S\geq s+1)$ for $s \in \{s \in \mathbb{N} | s\geq\underline{s}\}$, where $P(S\geq s)$ is computed from the respective continuous distribution.\footnote{Consequently, the probabilities add up to one by construction as $\sum_{s=\underline{s}}^{\infty} P(S=s) = P(S\geq \underline{s})=1$.} This discretization is adopted by  \citet{kondo_etal2023heavy} and advocated, for the Pareto case, by  \citet{buddana_kozubowski2014discrete}. Differently,  \citet{clauset_etal2009power} consider a power law for the probability mass function assuming $P(S=s) \equiv \zeta(k,\underline{s}) S^{-k-1}$, where $\zeta$ is a generalized zeta function, ``which is rather inconvenient to work within an applied setting'' \citep{buddana_kozubowski2014discrete}. 
	
	We use two estimators for the Pareto distribution. First, we apply Ordinary Least Squares (OLS) to Equation \eqref{eq:sf_pareto}, when we replace the survival function $P(S\geq s)$ by its empirical counterpart $\hat{P}(S\geq s)$, computed as the ratio between the number of firms with size $S\geq s$ and the number of firms with size $S\geq\underline{s}$.\footnote{Alternatively, we could apply OLS to a log-transformed histogram, gauging the power law exponent $k$ from the empirical probability function instead of the empirical survival function. We choose not to follow this strategy because this estimator performed very poorly in Monte Carlo simulations \citep{clauset_etal2009power,virkar_clauset2014power,bottazzi_etal2015zipf}.} Formally, given $\exp (\epsilon) \equiv \frac{\hat{P}(S\geq s)}{P(S\geq s)}$, the regression equation is
	\begin{align} 
		\ln \hat{P}(S\geq s) = & k \ln (\underline{s}/s)  + \epsilon \label{eq:ols_pareto}
	\end{align}	
	where $k$ is the only unknown parameter. Therefore, we do not follow the usual practice in the literature of freely estimating an intercept. By doing that, we address the concerns of \citet{clauset_etal2009power} that regression lines are not valid distributions since, in our approach, $P(S\geq \underline{s})=1$, which is not generally valid if an intercept is freely estimated.\footnote{ \citet{urzua2011testing} expresses similar concerns, arguing ``the intercept is not a nuisance parameter in the regression.''} Besides this intercept restriction, this method is essentially a standard rank-size regression with binned data, as the number of firms with size $S\geq s$ equals the rank size of a firm with exactly $s$ employees.\footnote{After all, if the $j$-th largest firm has size $s$, there must be $j$ firms with size $S\geq s$ if we assign the highest possible rank to firms with the same size (e.g., if the two largest firms are the same size, we assign rank 2 for both).} 
	
	Second, we use a maximum likelihood (ML) estimator.  \citet{virkar_clauset2014power} show that an analytical solution for this ML estimator (MLE) can be obtained when the binning scheme is logarithmic. For arbitrary bins such as those of Table \ref{tab:descrip}, however, a closed-form expression for this MLE does not exist, and thus, we obtain it numerically. Since it is computationally faster, we choose to solve the associated First-Order Condition (FOC), derived in  \ref{sec:app_mle_pareto}, instead of directly maximizing the log-likelihood function as in  \citet{virkar_clauset2014power}.
	
	If the correct $\underline{s}$ is chosen, it is known that OLS regression \eqref{eq:ols_pareto} consistently estimates $k$, since $\hat{P}(S\geq s)$ is a consistent estimator of $P(S\geq s)$ by the law of large numbers. It is also possible to show that MLEs for both binned and non-binned data are consistent and asymptotically efficient \citep{virkar_clauset2014power,clauset_etal2009power}.\footnote{The MLE for non-binned data is the known \citet{hill1975simple} estimator.} But how do they perform in small samples? This question is explored through Monte Carlo simulations by \citet{clauset_etal2009power}, \citet{virkar_clauset2014power}, and \citet{bottazzi_etal2015zipf}. They find OLS regression \eqref{eq:ols_pareto}, but without the intercept constraint, is biased in small samples, although this bias is not typically very high.\footnote{With the (correct) intercept constraint, one should expect a more efficient estimation of $k$. See \citet{schluter2018top} for proof of the rank-size regression case in large samples.} MLEs have the best performance in binned data \citep{virkar_clauset2014power} and also in non-binned data \citep{clauset_etal2009power,bottazzi_etal2015zipf}, accurately estimating $k$, with negligible bias.\footnote{For non-binned data, \citet{bottazzi_etal2015zipf} also find very good performance for the OLS rank-size estimator with Gabaix and Ibragimov's correction \citep{gabaix_ibragimov2011rank} .} These results are not unexpected as \citet{aban_meerschaert2004generalized} show that the MLE for non-binned data, with a small sample correction, is the best linear unbiased estimator (BLUE) and also the minimum variance unbiased estimator (MVUE).
	
	Finally, regarding the lognormal distribution, we follow  \citet{virkar_clauset2014power} and estimate its parameters $\mu$ and $\sigma>0$ using only the MLE for binned data. As in the Pareto case, there is no analytic expression for this estimator, and thus, we obtain it by numerically solving the FOCs for the likelihood maximization. See  \ref{sec:app_mle_lognormal} for the derivation of these FOCs.
	
	\subsection{Step 2: goodness-of-fit tests} \label{sec:method_step2}
	 \citet{virkar_clauset2014power} use a goodness-of-fit test to verify if a random variable follows an estimated distribution. This test requires a measure of the distance between empirical and estimated distributions. They suggest the Kolmogorov–Smirnov (KS) goodness-of-fit statistic, which can be formally defined as
	\begin{align} \label{eq:KS}
		D = & \max_{s \in \{\underline{s},...,500\}} \left|\hat{P}(S< s) - P(S<  s|\hat{\beta})\right| = \max_{s \in \{\underline{s},...,500\}} \left| \hat{P}(S\geq s) - P(S\geq s|\hat{\beta})\right|
	\end{align}
	where $\hat{P}(\cdot)$ is the empirical probability and $P(\cdot|\hat{\beta})$ is the probability under an evaluated distribution with the estimated vector of parameters $\hat{\beta}$. Given the distance measure \eqref{eq:KS}, an estimated distribution, and being $n$ the number of firms with at least $\underline{s}$ employees, the $p$-value of the test can be computed following five steps:
	\begin{enumerate}
		\item Compute the distance $D^*$ between estimated and empirical distributions using \eqref{eq:KS}. 	 	
		\item Generate a synthetic binned data set with $n$ values that follows the same estimated distribution above $\underline{s}$.
		\item Fit the model to this synthetic data set, obtaining a new estimated distribution.
		\item From \eqref{eq:KS}, compute the distance $D$ between this new model and the synthetic data set.
		\item Repeat steps 2–4 many times and report the fraction of the distances $D$ that are at least as large as $D^*$.
	\end{enumerate}	
	Some comments are due. First, in the second step of this algorithm,  \citet{virkar_clauset2014power} suggest the use of a semi-parametric bootstrap to generate a distribution that follows the estimated distribution above $\underline{s}$ and the empirical distribution below $\underline{s}$, which is necessary to them as they are also estimating $\underline{s}$. Since we are exogenously setting $\underline{s}$, we only need the distribution above $\underline{s}$. Second, they generate synthetic data above $\underline{s}$ by sampling from a non-binned distribution and then computing the synthetic bin counts. We choose to sample directly from a multinomial distribution whose events' probabilities are given by the probabilities of the bins, which can be easily computed from the estimated survival functions (see \ref{sec:app_mle}). Third, we generate 10,000 synthetic data sets for each test, which is probably high enough as \citet{virkar_clauset2014power} show that with 2,500 simulations, one can gauge the $p$-value to within 0.01 of the true value. Fourth, we compute the test for Pareto and lognormal distributions, for each considered estimator. We also test a strong Zipf's law, when no estimation is required as it is a Pareto distribution with $k=1$.
	
	\subsection{Step 3: comparing the distributions} \label{sec:method_step3}
	
	\citet{virkar_clauset2014power} suggest the use of the likelihood ratio test proposed by \citet{vuong1989likelihood} to compare non-nested distributions in binned data. Suppose one wants to compare distribution A against distribution B, which are not nested. Let $\mathcal{L}_d = \prod_{i=j}^m (p_{d,i})^{h_i}$ be the likelihood of distribution $d=A,B$, where $p_{d,i}$ is the probability that some observation falls within the $i$-th bin under distribution $d$ and $h_i$ is the number of raw observations in the $i$-th bin. Note that there are $m$ bins, but the distributions hold only from the $j$-th bin, meaning $\underline{s}$ is the lower bound of the $j$-th bin. Given that, the log-likelihood ratio of comparing A against B is $\mathcal{R} \equiv \ln\mathcal{L}_A - \ln \mathcal{L}_B$. Let us also define the normalized log-likelihood ratio as $\mathcal{R}_n \equiv \mathcal{R}/\sqrt{2n \hat{\sigma}_\mathcal{R}^2}$, where $\hat{\sigma}_\mathcal{R}^2$ is the estimated variance on the log-likelihood ratio $\mathcal{R}$, that is,
	\begin{align} \label{eq:sigma2_vuong}
		\hat{\sigma}_\mathcal{R}^2 \equiv & \frac{1}{n} \sum_{i=j}^m h_i \left[\left(\ln p_{A,i}-\ln p_{B,i}\right) - \mathcal{R}/n \right]^2
	\end{align}
	$n \equiv \sum_{i=j}^{m} h_i$ is the number of firms with at least $\underline{s}$ employees or, equivalently, the number of firms at the $j$-th bin or above.  \citet{vuong1989likelihood} shows that under the null that the two distributions are equivalent, $\sqrt{2} \mathcal{R}_n \xrightarrow{D} N(0,1)$; under the alternative that distribution A is better, $\sqrt{2} \mathcal{R}_n \xrightarrow{a.s.} +\infty$; finally, under the alternative that distribution B is better, $\sqrt{2} \mathcal{R}_n \xrightarrow{a.s.} -\infty$. As a consequence, under the null hypothesis, in large samples,
	\begin{align} \label{eq:p_vuong}
		P\left(|\mathcal{R}|\geq|\mathcal{R}^*|\right) = & P\left(\sqrt{2}|\mathcal{R}_n|\geq\sqrt{2}|\mathcal{R}_n^*|\right) = 2 \times P\left(\sqrt{2}\mathcal{R}_n\geq\sqrt{2}|\mathcal{R}_n^*|\right) \notag \\
		P\left(|\mathcal{R}|\geq|\mathcal{R}^*|\right) = & 2 \left\{1 - (1/2)\left[1+\erf\left(\sqrt{2}|\mathcal{R}_n^*|/\sqrt{2}\right) \right] \right\} = 1-\erf\left(|\mathcal{R}_n^*|\right)
	\end{align}
	where $\erf(z) \equiv \frac{2}{\sqrt{\pi}}\int_0^z e^{-t^2}dt$ is the Gaussian error function. Hence, setting a significance level $p^*$, one can get $T>0$ that solves $p^* = 1- \erf(T)$. If $\mathcal{R}_n \geq T$ ($\mathcal{R}_n \leq -T$), the null is rejected in favor of A being better (worse) than B, while the null is not rejected if $-T < \mathcal{R}_n < T $.
	
	We apply this test to compare (i) Pareto against lognormal and (ii) strong Zipf's law against lognormal, using Pareto and lognormal densities as estimated by ML. Testing strong Zipf's law against the Pareto distribution is equivalent to verifying if $k=1$. However, standard OLS t-tests would not be reliable here since they have a strong tendency to over-reject the null $k=1$, as  \citet{gabaix_ibragimov2011rank} and  \citet{bottazzi_etal2015zipf} show through Monte Carlo exercises. Indeed, when sampling from a Zipf distribution, \citet{bottazzi_etal2015zipf} could reject the null $k=1$ at 5\% confidence level 60 -- 70\% of the time! Given that, we follow  \citet{virkar_clauset2014power} and verify it using the ML estimates and a standard likelihood ratio test. Under the null $k=1$, it is known that $2|\mathcal{R}|$ is asymptotically chi-squared distributed with one degree of freedom, where $\mathcal{R}$ is the log-likelihood ratio between Pareto and Zipf distributions.
	
	\section{Results} \label{sec:results}
	
	\subsection{Step 1: fitting the distributions} \label{sec:results_step1}
	
\begin{figure}[p]
		\centering
		\includegraphics[scale=0.87]{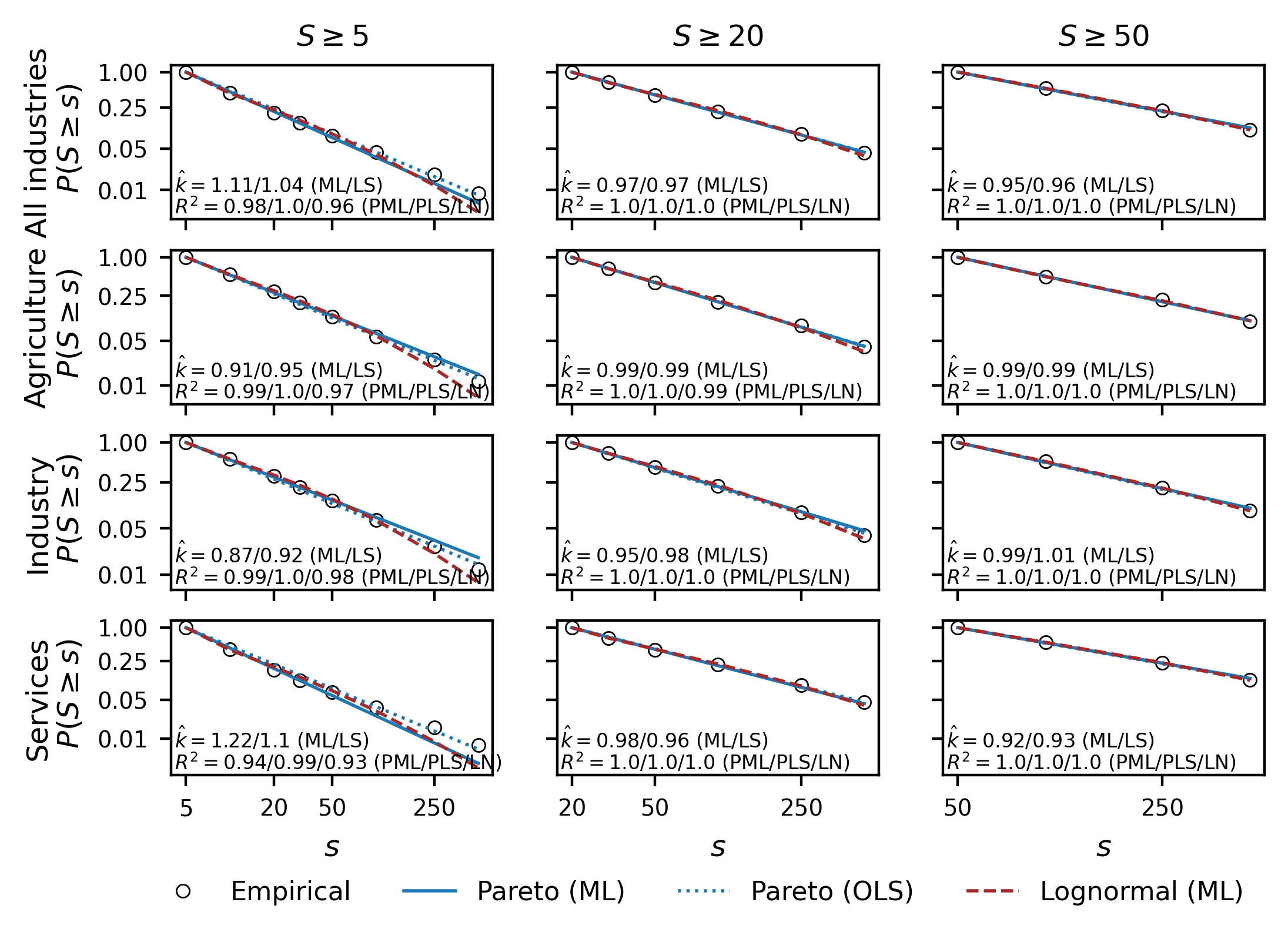}
		\caption{Models fit in 1996 (axes in logarithmic scale).}
		\label{fig:1996_plot}
	\end{figure}

\begin{figure}[p]
		\centering
		\includegraphics[scale=0.87]{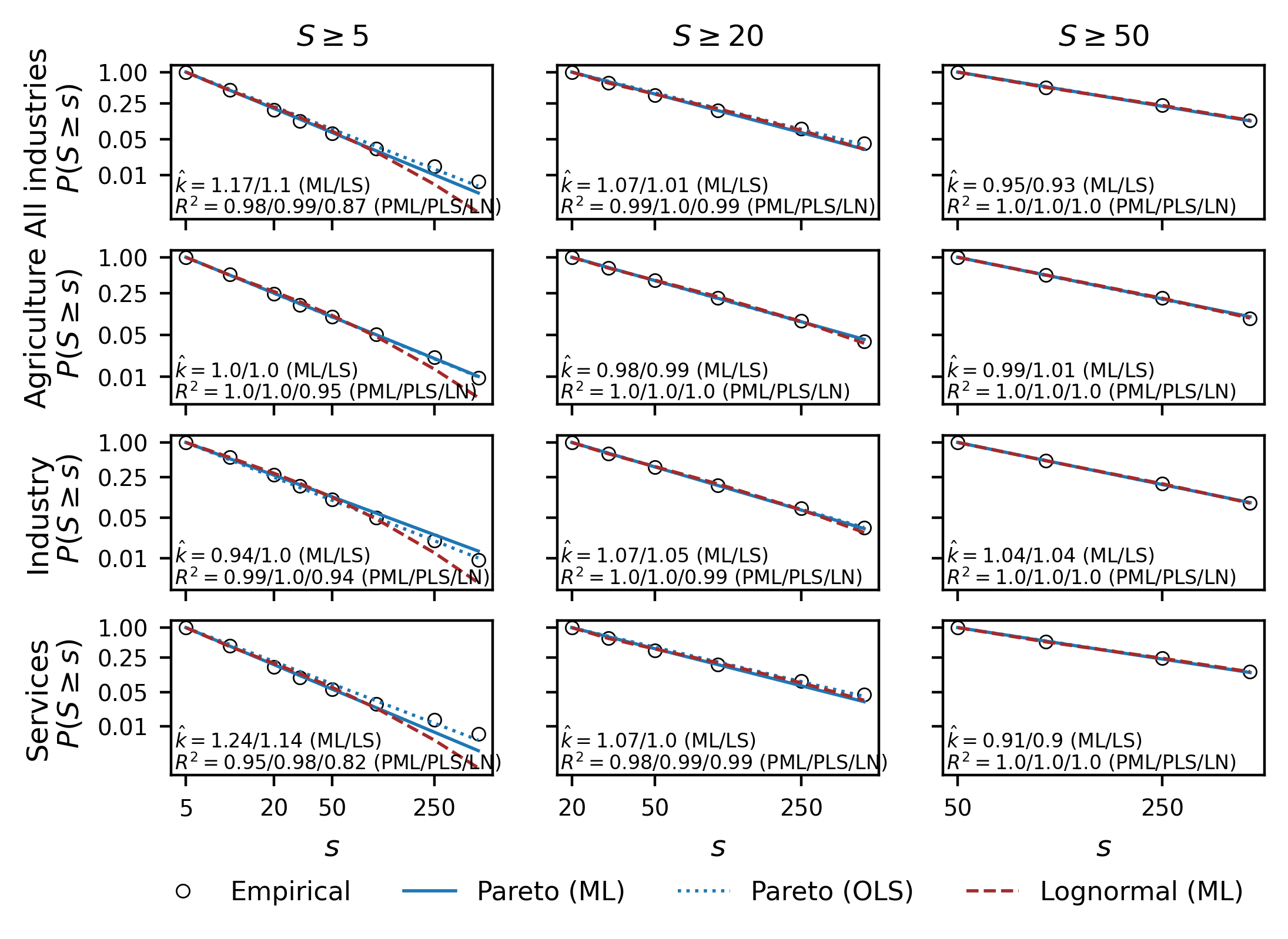}
		\caption{Models fit in 2020 (axes in logarithmic scale).}
		\label{fig:2020_plot}
	\end{figure}

	Figures \ref{fig:1996_plot} and \ref{fig:2020_plot} plot empirical and estimated survival functions in 1996 and 2020, respectively, our sample’s initial and final years. In both cases, we present the results at economy-wide and industry levels, with $\underline{s}=5,20,50$. The axes of each plot are in logarithmic scale, with $P(S\geq s)$ in the vertical axis and $s$ in the horizontal axis, implying estimated survival functions are straight lines in Pareto cases. Inside each plot, we show the OLS/ML estimates of $k$, $\hat{k}$, and the (centered) $R^2$ for each estimator/distribution computed from these plotted data. As can be seen, both distributions fit the data well. For $\underline{s}=5$, the Pareto distribution does a better job, especially closer to the upper tail, while for $\underline{s}=20$ and mainly for $\underline{s}=50$, both distributions fit similarly well. Focusing on the Pareto case, note estimates of $k$ are relatively robust to the choice of estimator, particularly for higher $\underline{s}$. Additionally, all estimates of $k$ are around one and typically become closer to this level as $\underline{s}$ increases. 
	
\begin{figure}[p]
		\centering
		\includegraphics[scale=0.87]{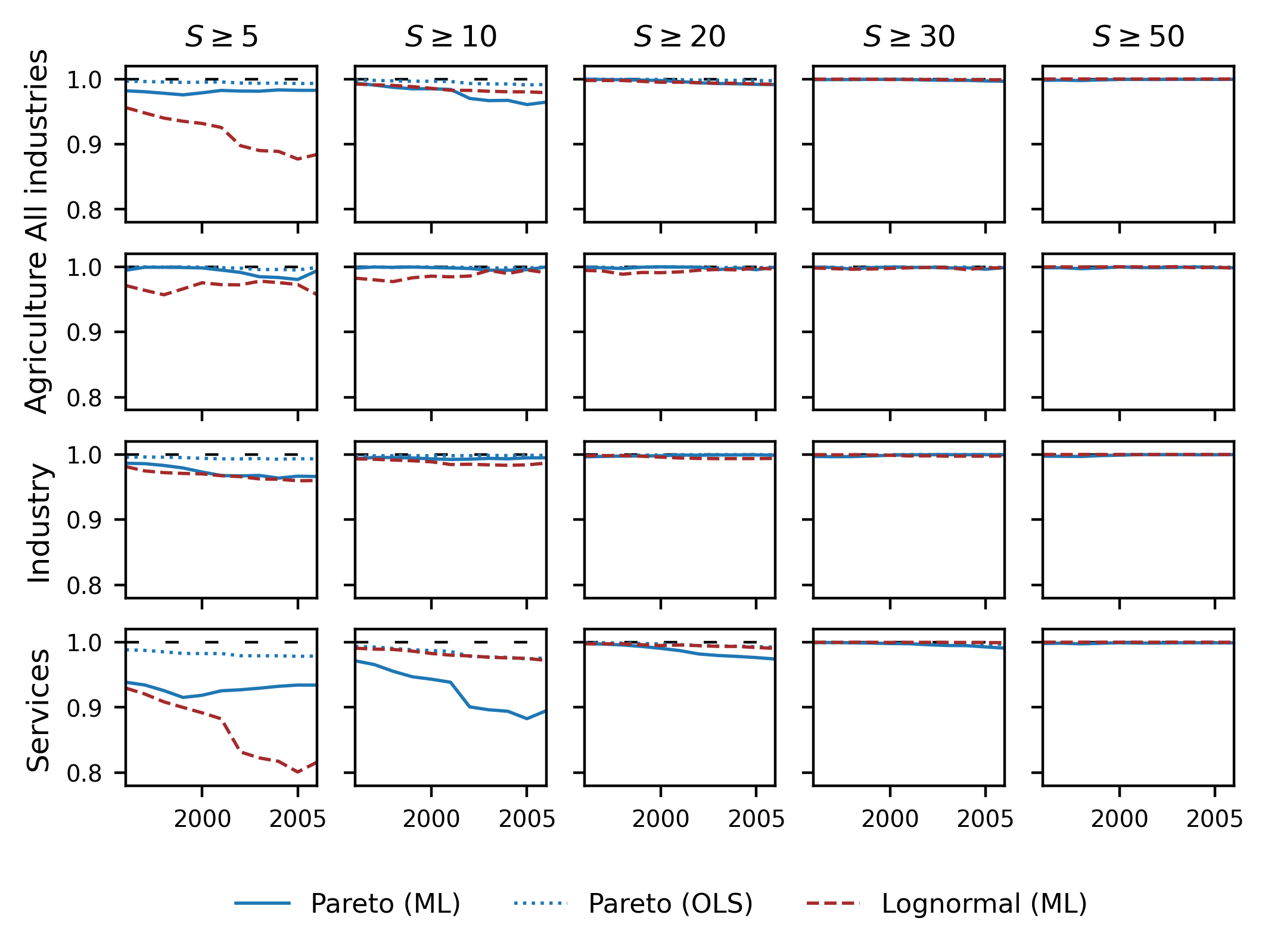}
		\caption{Centered $R^2$, 1996-2006.}
		\label{fig:R2_1996-2006}
	\end{figure}

\begin{figure}[p]
		\centering
		\includegraphics[scale=0.87]{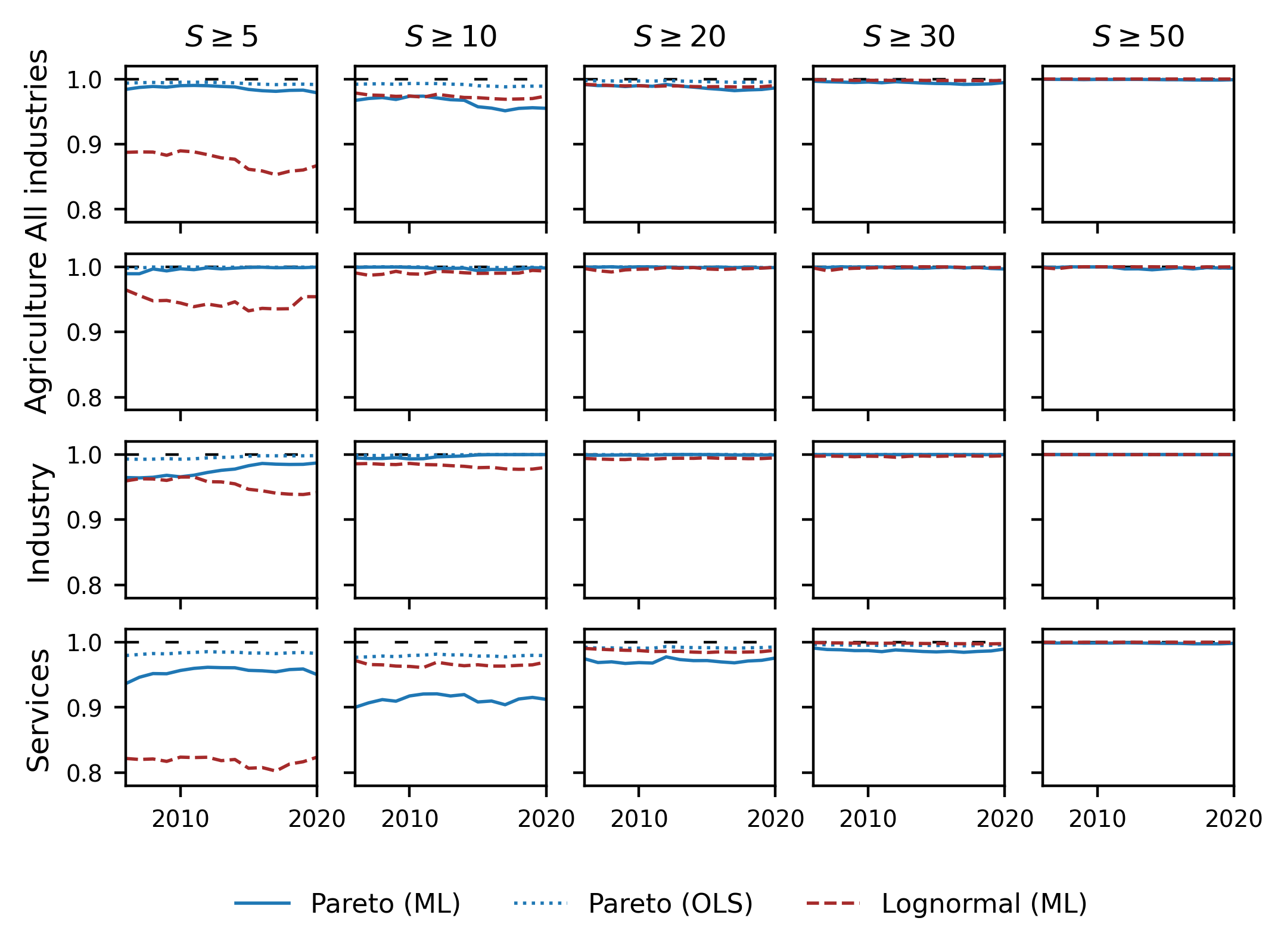}
		\caption{Centered $R^2$, 2006-2020.}
		\label{fig:R2_2006-2020}
	\end{figure}

\begin{figure}[p]
		\centering
		\includegraphics[scale=0.87]{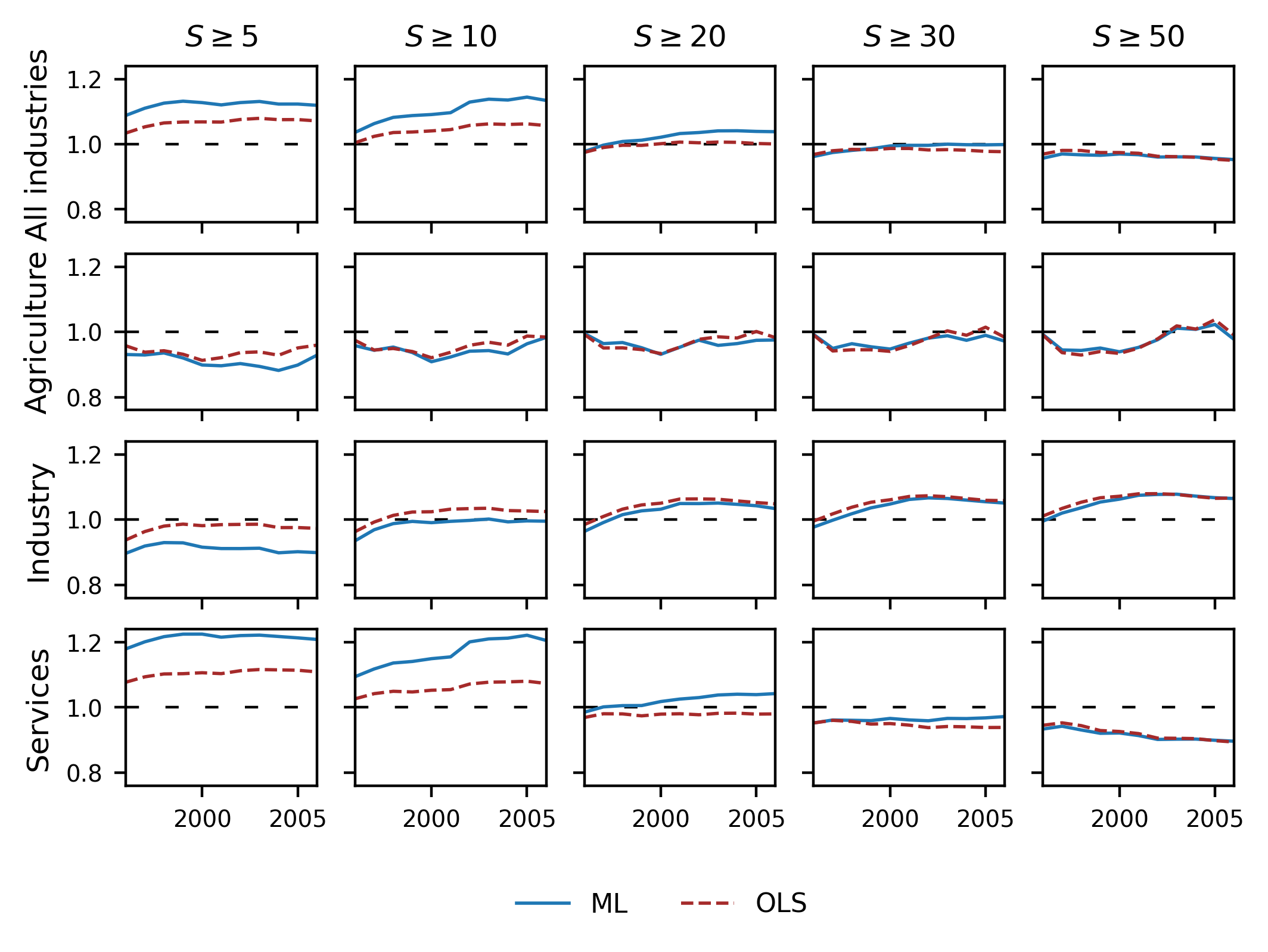}
		\caption{$k$ estimates, 1996-2006.}
		\label{fig:k_1996-2006}
	\end{figure}

\begin{figure}[p]
		\centering
		\includegraphics[scale=0.87]{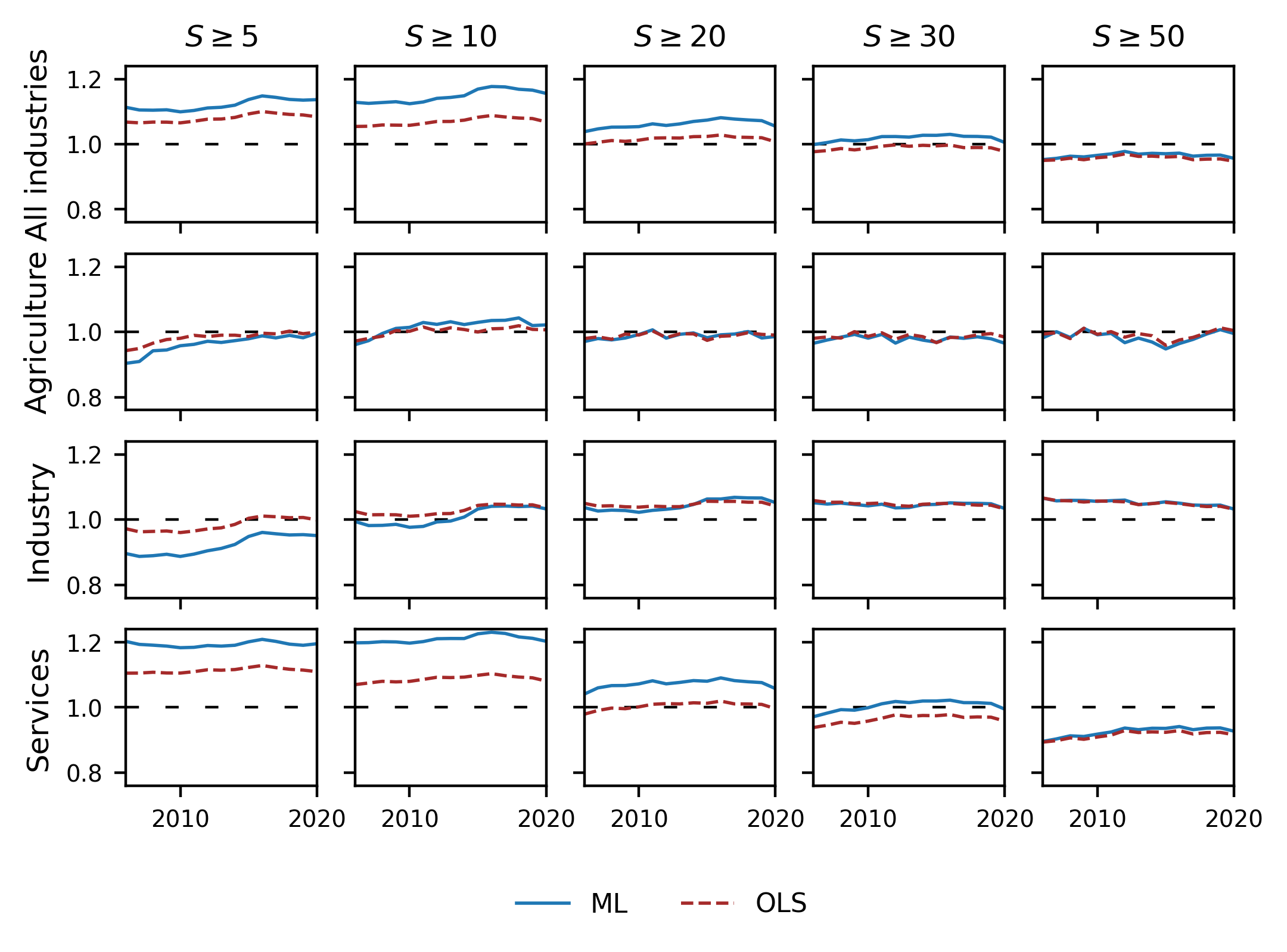}
		\caption{$k$ estimates, 2006-2020.}
		\label{fig:k_2006-2020}
	\end{figure}

	These results are not specific to 1996 and 2020 or $\underline{s}=5,20,50$. In Figures \ref{fig:R2_1996-2006} and \ref{fig:R2_2006-2020}, we plot the (centered) $R^2$ for each year, industry, lower bound $\underline{s}$, and estimator/distribution for 1996-2006 and 2006-2020, respectively. The fit of each model is very good for $\underline{s}=20,30,50$, while for $\underline{s}=5$, the lognormal fit is usually worse. Moreover, especially for $\underline{s}=10,20$, the ML Pareto estimate has the worst fit for the services sector. The power law exponent $k$ estimates for 1996-2006 and 2006-2020 are shown in Figures \ref{fig:k_1996-2006} and \ref{fig:k_2006-2020}, respectively. Several things are worth noting about these estimates. First, they are around one, typically between 0.8 and 1.2, and approach the unitary value for higher values of $\underline{s}$. Second, they are also surprisingly stable over time.  \citet{fujiwara_etal2004pareto} find similar stability for the UK, France, Italy, and Spain between 1993 and 2001, with firm size measured by total assets, number of employees, and sales (except for the UK).  \citet{resende_cardoso2022firm}, using net revenue to measure firm size, also estimate a relatively stable power law exponent for Brazil between 1999 and 2019. Third, OLS estimates vary much less than those by ML when a different $\underline{s}$ is chosen. This finding aligns with the results of \citet{aban_meerschaert2004generalized} for the daily trading volume of Amazon, Inc. stock, and \citet{kratz_resnick1996qq} in both empirical and Monte Carlo exercises.\footnote{Their Monte Carlo experiment uses a Pareto with $k=1$, while they empirically assess ``[...] interarrival times between packets generated and sent to a host by a terminal during	a logged-on session'' \citep{kratz_resnick1996qq}.}

	\subsection{Step 2: goodness-of-fit tests} \label{sec:results_step2}
	
\begin{figure}[p]
		\centering
		\includegraphics[scale=0.87]{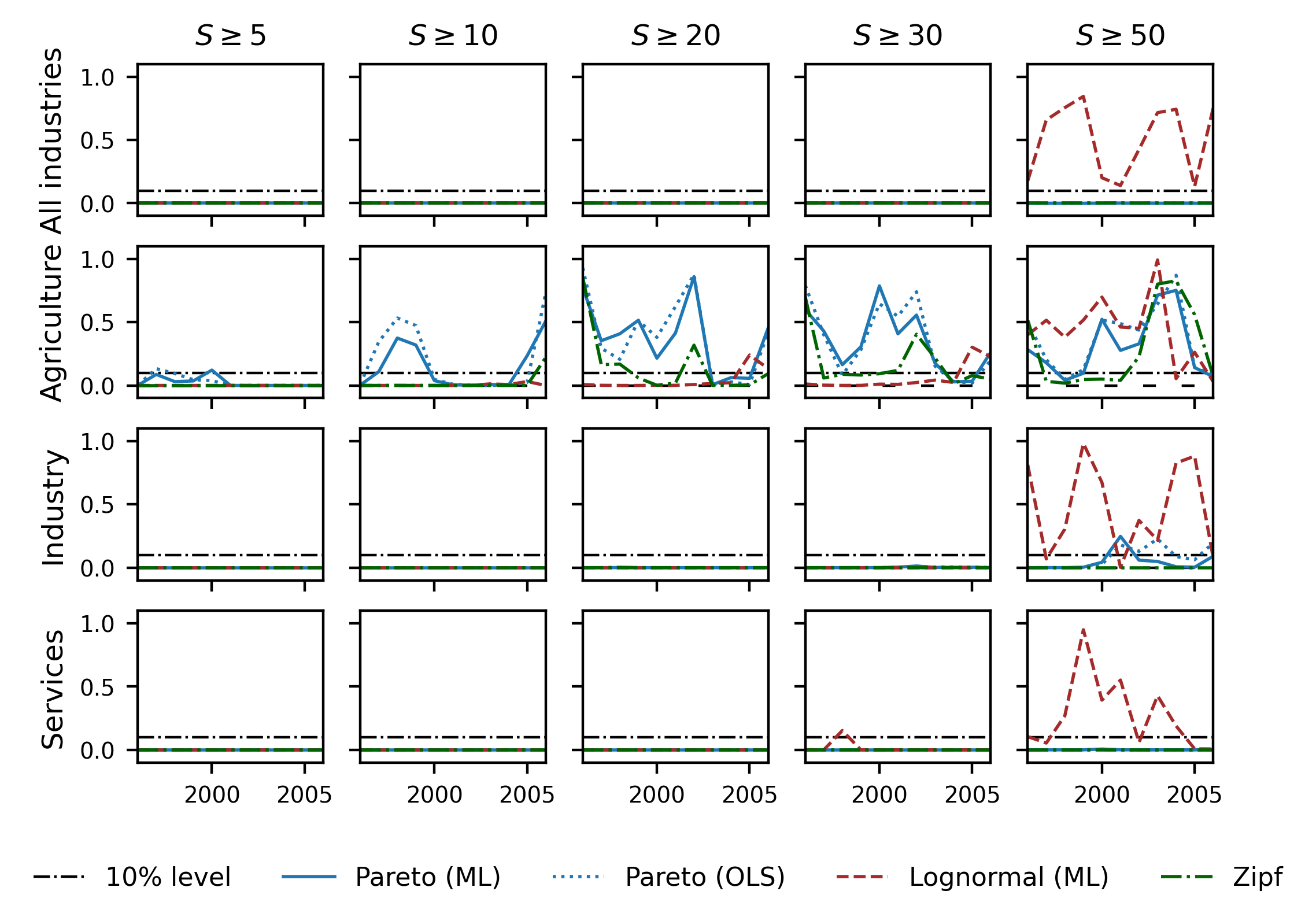}
		\caption{$p$-value of the goodness-of-fit test, 1996-2006.}
		\label{fig:fit_test_1996-2006}
	\end{figure}

\begin{figure}[p]
		\centering
		\includegraphics[scale=0.87]{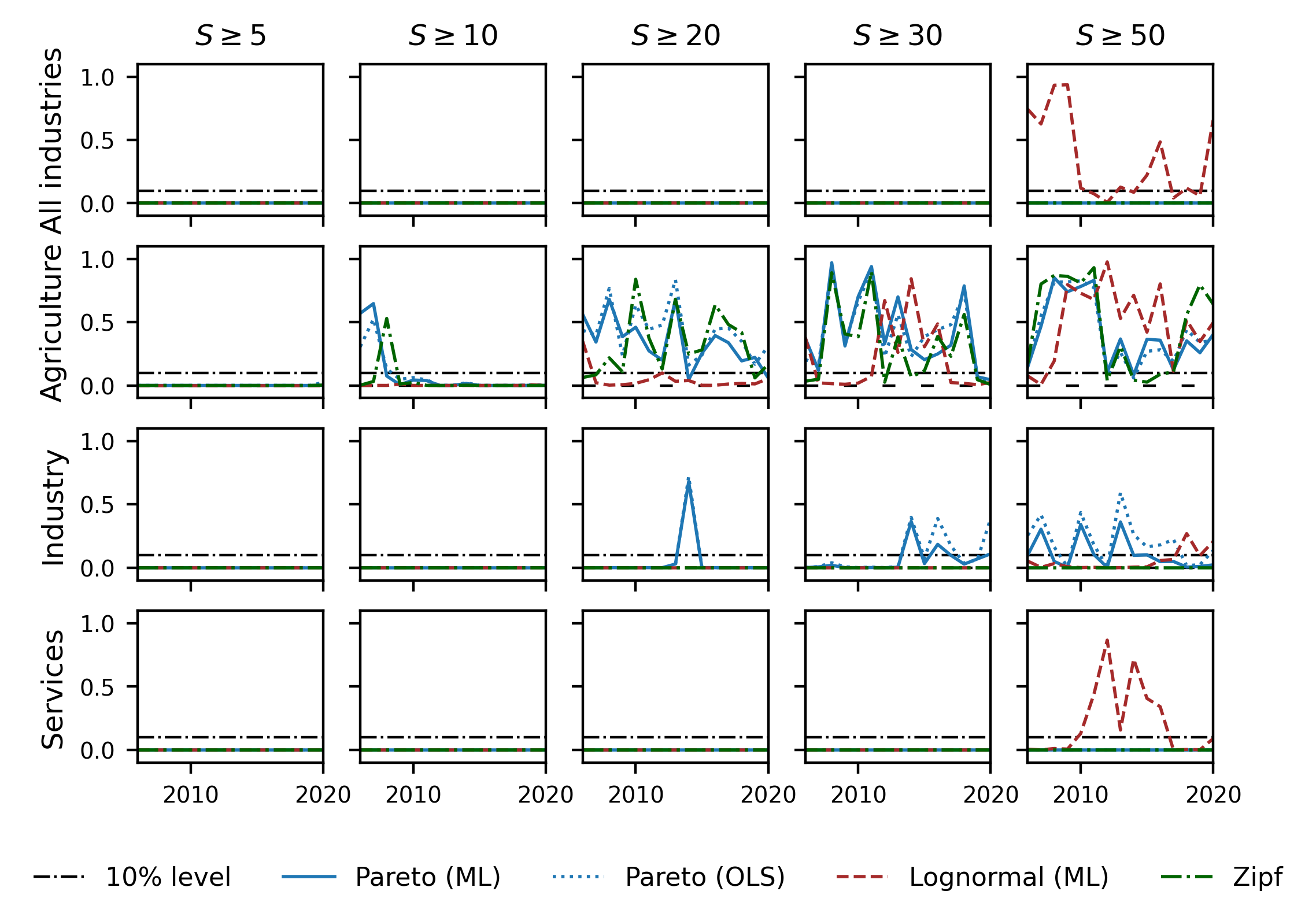}
		\caption{$p$-value of the goodness-of-fit test, 2006-2020.}
		\label{fig:fit_test_2006-2020}
	\end{figure}

	The computed $p$-values for the goodness-of-fit tests are shown in Figures \ref{fig:fit_test_1996-2006} and \ref{fig:fit_test_2006-2020} for 1996-2006 and 2006-2020, respectively, which are essentially the same for ML and OLS estimators of the Pareto density. Besides the very good fit of the estimated distributions shown previously, these tests reject both Pareto and lognormal distributions in most cases. Consistent with these findings, \citet{resende2004lei} does not find strong evidence supporting a lognormal distribution of firm size by the number of employees in Brazil either. Similarly, the Zipf distribution is also usually rejected. For Pareto and Zipf distributions, the main exception to these conclusions is agriculture, particularly for $\underline{s} = 20,30,50$ when the Pareto distribution is not rejected in almost all years, and the strong Zipf's law cannot be rejected for several years between 2006 and 2020. In the lognormal case, the main exception is $\underline{s} = 50$, when the distribution is typically not rejected (except for industry between 2006 and 2020).
	
	\subsection{Step 3: comparing the distributions} \label{sec:results_step3}
	
\begin{figure}[p]
		\centering
		\includegraphics[scale=0.87]{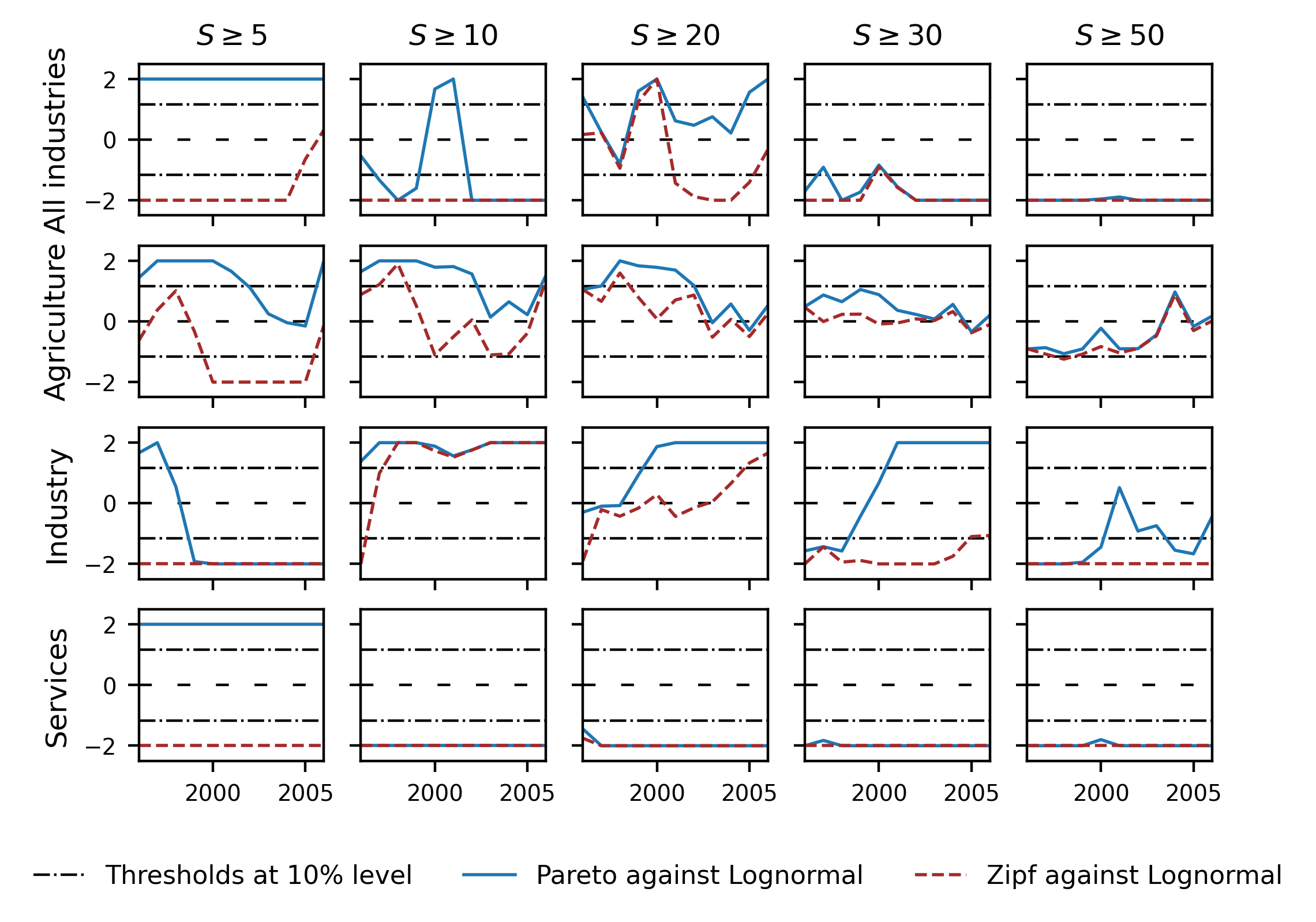}
		\caption{Normalized log-likelihood ratio, 1996-2006.}
		\label{fig:Vuong_test_1996-2006}
	\end{figure}

\begin{figure}[p]
		\centering
		\includegraphics[scale=0.87]{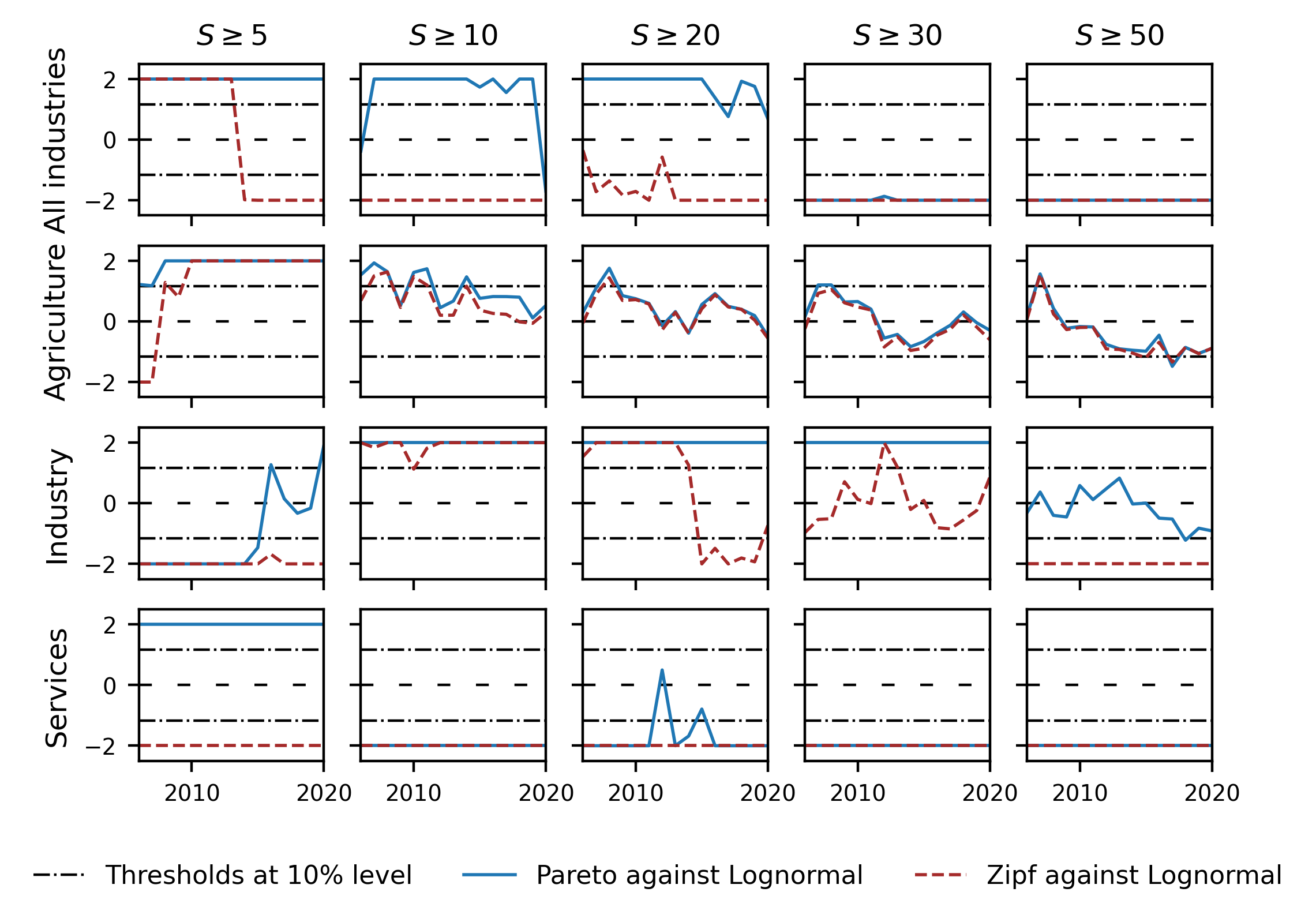}
		\caption{Normalized log-likelihood ratio, 2006-2020.}
		\label{fig:Vuong_test_2006-2020}
	\end{figure}

	In Figures \ref{fig:Vuong_test_1996-2006} and \ref{fig:Vuong_test_2006-2020}, we plot the normalized log-likelihood ratio $\mathcal{R}_n$ and the thresholds at 10\% level for 1996-2006 and 2006-2020, respectively. To make it easier to visualize the results, we plot $\mathcal{R}_n = 2$ ($\mathcal{R}_n = -2$) when $\mathcal{R}_n \geq 2$ ($\mathcal{R}_n \leq -2$). The results confirm that the lognormal provides a strong test for the Pareto distribution since there is no single winner between them in all cases. Typically, the Pareto distribution beats the lognormal for lower $\underline{s}$, while the lognormal wins for higher $\underline{s}$, particularly for $\underline{s}=50$, which is consistent with the goodness-of-fit tests results seen in the last section. Furthermore, it is worth mentioning that several of these results are consistent with the $R^2$ shown in Figures \ref{fig:R2_1996-2006} and \ref{fig:R2_2006-2020}. For instance, both likelihood and $R^2$ of the Pareto distribution are mostly higher for $\underline{s}=5$ but lower in the services sector for $\underline{s}=10,20,30,50$. Finally, when comparing strong Zipf's law and lognormal, the latter rarely loses. The main exception is industry under $\underline{s}=10$.		
	
\begin{figure}[p]
		\centering
		\includegraphics[scale=0.87]{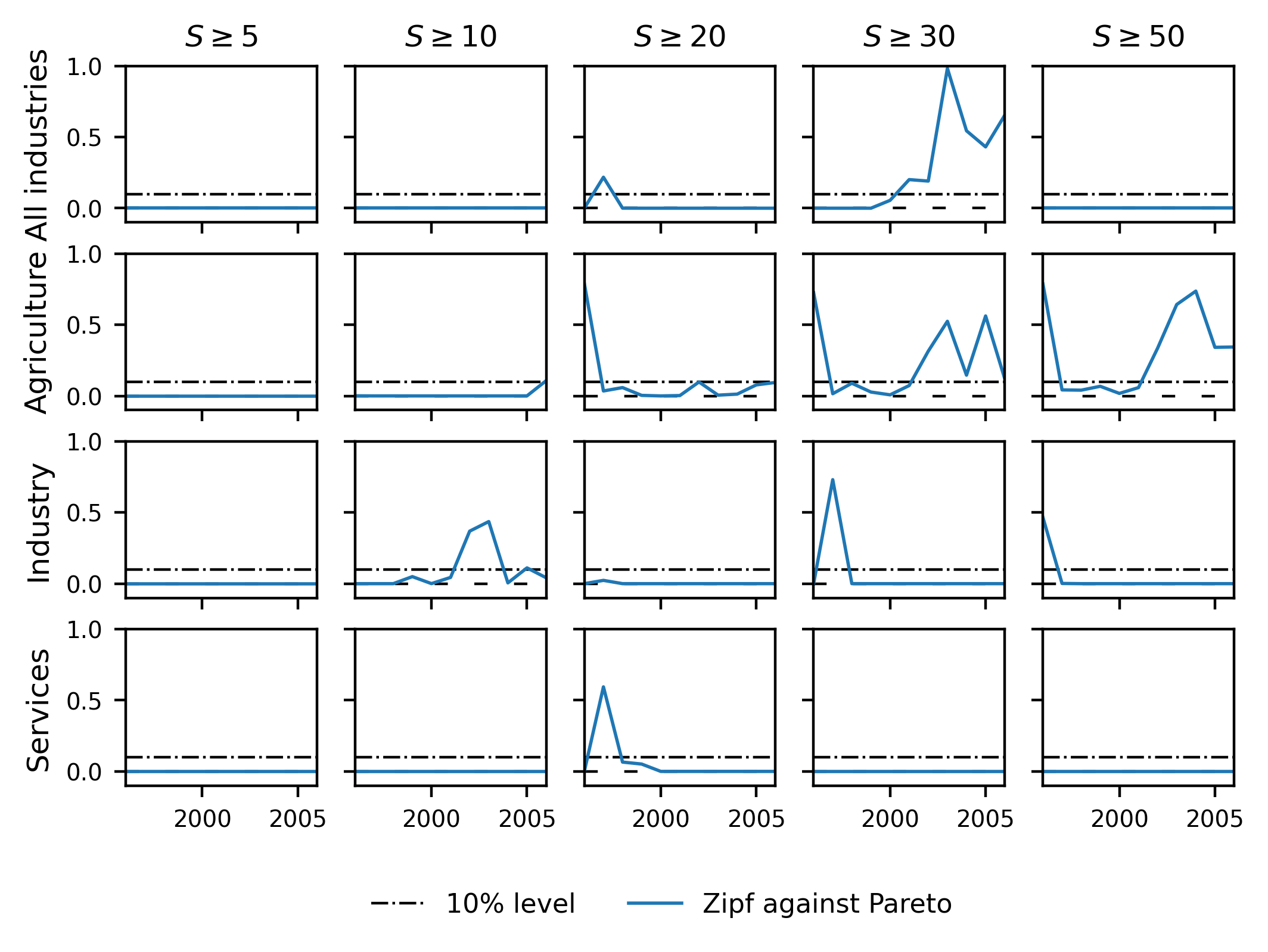}
		\caption{$p$-value of the standard likelihood ratio test, 1996-2006.}
		\label{fig:Zipf_test_1996-2006}
	\end{figure}

\begin{figure}[p]
		\centering
		\includegraphics[scale=0.87]{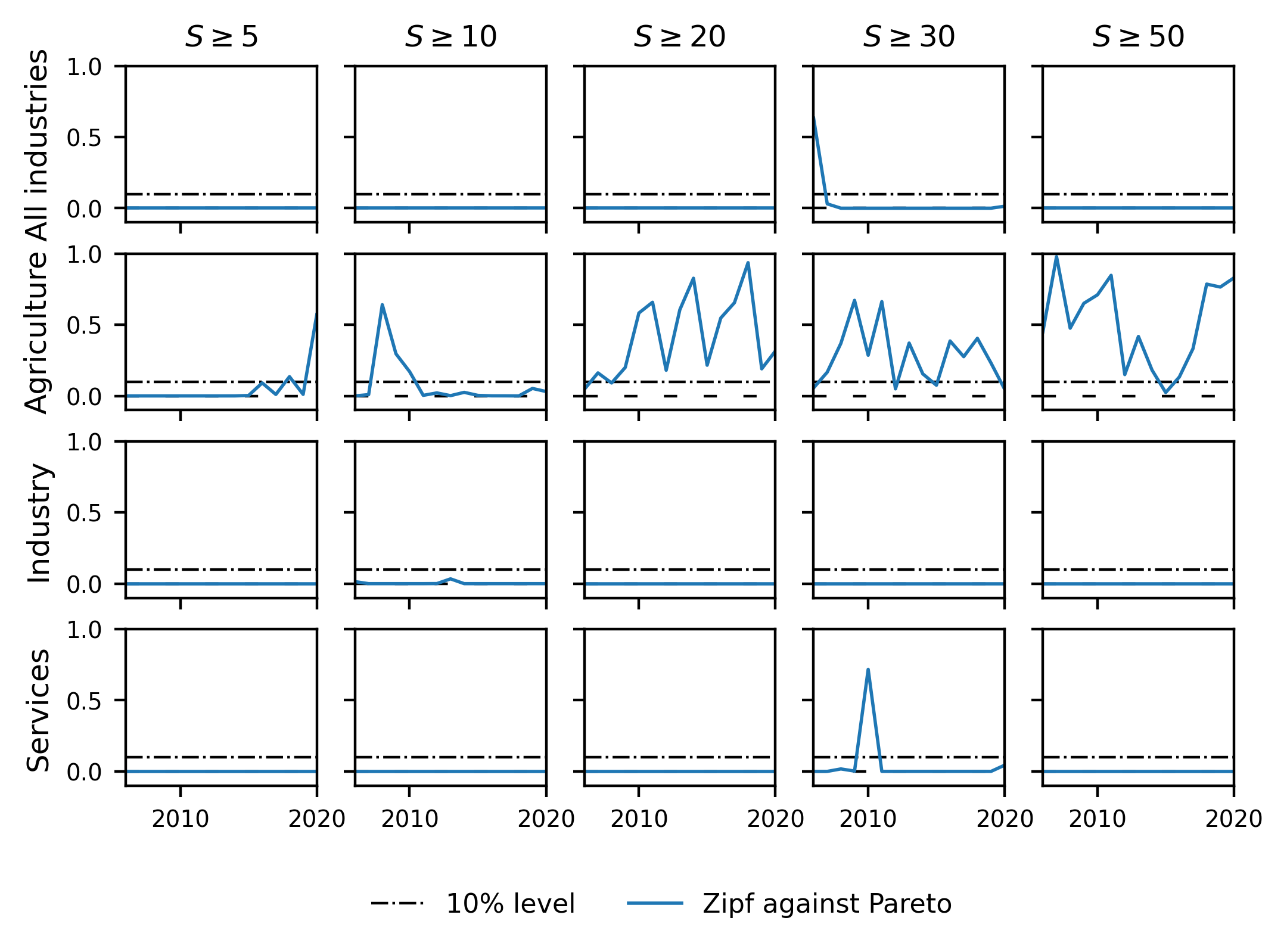}
		\caption{$p$-value of the standard likelihood ratio test, 2006-2020.}
		\label{fig:Zipf_test_2006-2020}
	\end{figure}

	The $p$-values of testing strong Zipf's law against the Pareto distribution for 1996-2006 and 2006-2020 are shown in Figures \ref{fig:Zipf_test_1996-2006} and \ref{fig:Zipf_test_2006-2020}, respectively. In almost all industries and years, we can reject $k=1$. The main exception is 2006-2020 agriculture under $\underline{s}=20,30,50$. Therefore, although the estimates of the power law exponent $k$ are around one, especially for higher $\underline{s}$, they are not \textit{exactly} one in most cases.
	
	\subsection{Discussion} \label{sec:results_discussion}	
\begin{figure}[h!]
		\centering
		\includegraphics[scale=1]{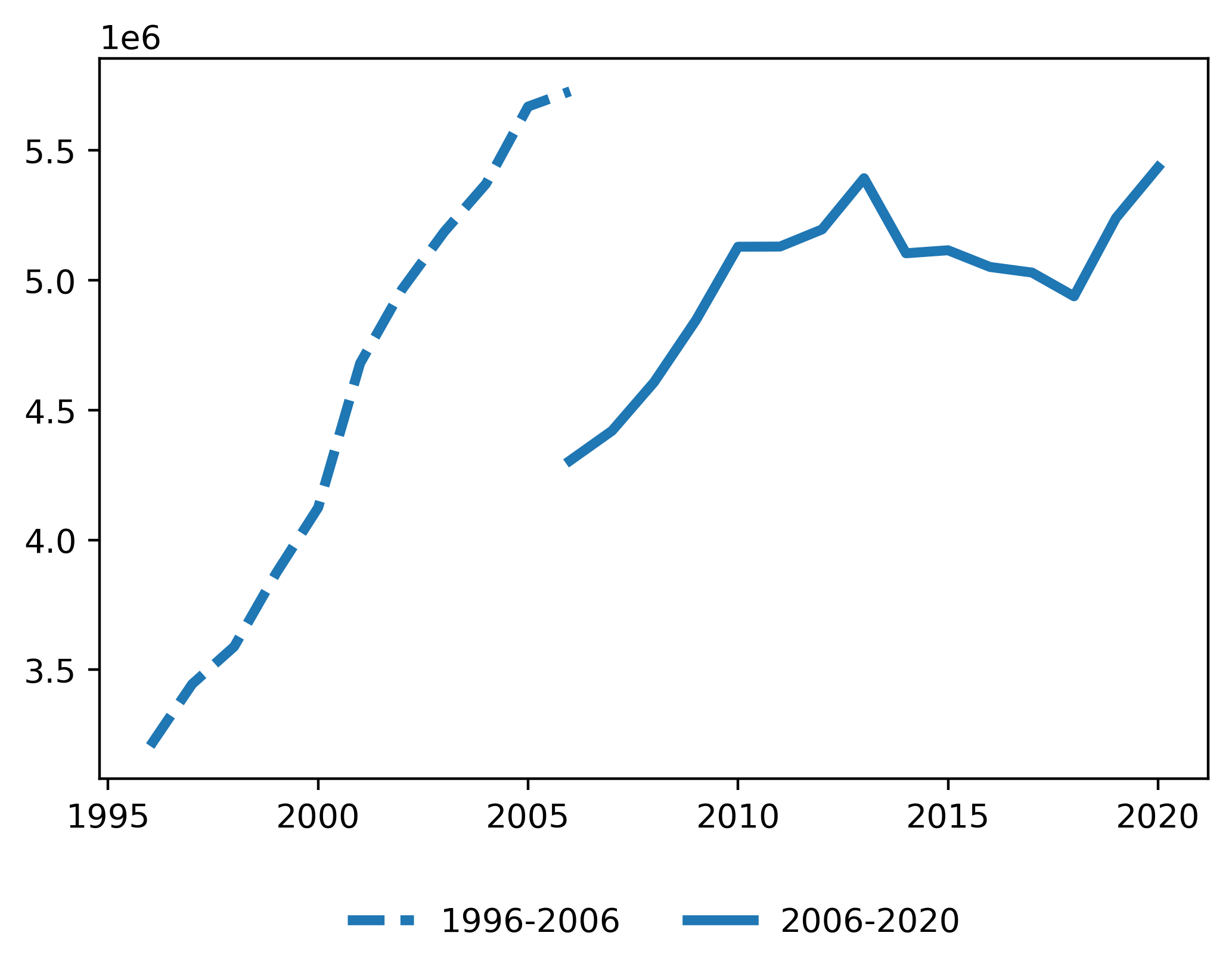}
		\caption{Total number of firms.}
		\label{fig:tot_firms}
	\end{figure}

	Let us summarize and discuss our findings from all three steps. Although a Zipf distribution can be ruled out, we estimate power exponent $k\approx1$ with good data fit, especially for higher $\underline{s}$, consistent with Zipf's law. However, a lognormal density also performs well and even outperforms the Pareto distribution in certain cases. The main issue is that the goodness-of-fit tests ruled out that the firm size distribution in Brazil is \textit{exactly} Pareto, Zipf, or lognormal in most cases. Nevertheless, as \citet{gabaix2009power} points out, \blockquote{With an infinitely large empirical data set, one can reject any nontautological theory. Hence, the main question of empirical work should be how well a theory fits, rather than whether it fits perfectly (i.e., within the standard errors). [...] Consistent with these suggestions, some of the debate on Zipf’s law should be cast in terms of how well, or poorly, it fits, rather than whether it can be rejected.} From that point of view, Pareto and lognormal distributions are still useful benchmarks as they provide very good, although not perfect, approximation to data. This can be seen more clearly in Table \ref{tab:P_B}, which shows empirical and estimated bins' probabilities over the support $S\geq20$ for ML Pareto and lognormal distributions in 1996 and 2020. These good fits hold at the economy-wide level and also for agriculture, industry, and services alone, for each year between 1996 and 2020. As it is well known, Brazil experienced an economic boom in the 2000s and a bust with huge volatility in the 2010s, which possibly explains why the total number of firms varied so much over time (Figure \ref{fig:tot_firms}), but firm size distribution remained basically unchanged throughout the entire period, always close to Zipf's law. This is a rather remarkable result even if this ``law'' is not \textit{exactly} valid, since, as \citet{gabaix2009power} points out for the distribution of city size, ``there is no tautology causing the data to automatically generate this shape.''
	
	\begin{table}[h!]
		\centering
		\caption{Empirical and ML estimated bins' probabilities over $S\geq20$}
		\label{tab:P_B}
		\begin{tabular}{lcccccccc}
			\hline \hline
			Number of & &  \multicolumn{3}{c}{1996} & & \multicolumn{3}{c}{2020} \\ \cline{3-5} \cline{7-9} 
			employees & & Empirical & Pareto & Lognormal & & Empirical & Pareto & Lognormal \\
			\hline	
			\multicolumn{2}{l}{\textit{All industries}} \\	
			20 to 29 &              &             33.6 &          32.6 &             34.1 &              &             38.0 &          35.2 &             38.9 \\
			30 to 49 &              &             25.8 &          26.4 &             24.2 &              &             26.5 &          27.3 &             23.8 \\
			50 to 99 &              &             19.1 &          20.1 &             19.4 &              &             17.6 &          19.7 &             17.9 \\
			100 to 249 &              &             12.6 &          12.3 &             13.6 &              &              9.9 &          11.2 &             12.0 \\
			250 to 499 &              &              4.7 &           4.2 &              4.9 &              &              4.0 &           3.5 &              4.2 \\
			500 or more &              &              4.3 &           4.4 &              3.8 &              &              4.1 &           3.2 &              3.2 \\
			\hline
			\multicolumn{2}{l}{\textit{Agriculture}}\textit{} \\
			20 to 29 &              &             33.4 &          33.2 &             34.0 &              &             34.2 &          32.9 &             34.3 \\
			30 to 49 &              &             26.5 &          26.6 &             24.9 &              &             24.4 &          26.5 &             24.4 \\
			50 to 99 &              &             20.3 &          20.0 &             19.7 &              &             20.5 &          20.1 &             19.4 \\
			100 to 249 &              &             11.2 &          12.1 &             13.4 &              &             12.2 &          12.2 &             13.4 \\
			250 to 499 &              &              4.7 &           4.0 &              4.7 &              &              4.8 &           4.1 &              4.8 \\
			500 or more &              &              4.0 &           4.1 &              3.4 &              &              3.9 &           4.2 &              3.7 \\
			\hline
			\multicolumn{2}{l}{\textit{Industry}}\textit{} \\
			20 to 29 &              &             31.2 &          32.1 &             31.6 &              &             36.1 &          35.1 &             36.6 \\
			30 to 49 &              &             26.2 &          26.2 &             25.2 &              &             26.4 &          27.3 &             25.1 \\
			50 to 99 &              &             20.7 &          20.2 &             20.6 &              &             19.3 &          19.7 &             19.1 \\
			100 to 249 &              &             13.2 &          12.5 &             14.2 &              &             10.9 &          11.2 &             12.3 \\
			250 to 499 &              &              4.8 &           4.3 &              4.9 &              &              3.8 &           3.5 &              4.1 \\
			500 or more &              &              3.9 &           4.6 &              3.5 &              &              3.4 &           3.2 &              2.8 \\
			\hline
			\multicolumn{2}{l}{\textit{Services}}\textit{} \\
			20 to 29 &              &             34.9 &          32.8 &             35.5 &              &             38.8 &          35.3 &             39.8 \\
			30 to 49 &              &             25.5 &          26.5 &             23.6 &              &             26.5 &          27.3 &             23.2 \\
			50 to 99 &              &             18.1 &          20.1 &             18.8 &              &             16.9 &          19.6 &             17.5 \\
			100 to 249 &              &             12.3 &          12.2 &             13.2 &              &              9.5 &          11.1 &             11.8 \\
			250 to 499 &              &              4.7 &           4.1 &              4.9 &              &              4.0 &           3.5 &              4.3 \\
			500 or more &              &              4.5 &           4.2 &              4.0 &              &              4.4 &           3.2 &              3.4 \\
			\hline \hline					
		\end{tabular}		
	\end{table}	
	
	\section{Conclusion} \label{sec:conclusion}
	
	In this paper, we evaluate Zipf's law for the distribution of firm size by the number of employees in Brazil. Remarkably, we find that Zipf's law provides a very good, although not perfect, approximation to data for each year between 1996 and 2020 at the economy-wide level and also for agriculture, industry, and services alone. However, a lognormal distribution also performs well and even outperforms Zipf's law in certain cases.
	
	Our analyses are based on publicly available data from CEMPRE, which facilitates other researchers' reproduction and exploration of our results. Nevertheless, this choice also has relevant shortcomings due to binning, suggesting working with CEMPRE firm-level data may be an interesting avenue for future research. First, binning leads to a loss of information, such that a higher number of sampled firms is required to achieve the same accuracy in estimating and testing the distributions when data is binned \citep{virkar_clauset2014power}. One may argue that this information loss could be especially harsh in the CEMPRE database since there is little information on the upper tail. After all, the last bin available contains firms with 500 or more employees, which is probably too wide since the biggest firms would typically have a much larger number of employees. In any case, since our samples are large, this may not be such a severe problem here. Second, one can easily explore more flexible distributions when working with non-binned data.  \citet{kondo_etal2023heavy} estimate statistical mixtures and convolutions of Pareto and lognormal distributions in the US, finding these combinations significantly beat each distribution alone. Alternatively, one can apply Lagrange multiplier tests, verifying the null of power or Zipf's law against a distribution that \textit{nests} the Pareto density. One advantage of these tests is that they do not require the estimation of the more general density, which may be challenging in some cases. In principle, such tests could be applied to binned data; however, to the best of our knowledge, so far, they have been developed only for non-binned data (see, e.g., \citet{urzua2000simple} for testing Zipf's law against Pareto type II, \citet{goerlich2013simple} for testing power law against Pareto type II, and \citet{urzua2020simple} for testing power law against Pareto type IV). \citet{resende_cardoso2022firm} apply these tests to the distribution of firm size by net revenue in Brazil. They consider the 1,000, 500, and 100 largest firms between 1999 and 2019, finding strong support for power or Zipf's law only (i) against the Pareto type II distribution and (ii) among the 100 largest firms. This suggests investigating distributions that nest the Pareto density using CEMPRE firm-level data can be fruitful.

\vspace{2.0cm}

\section*{Acknowledgements}

DOC thanks to CNPq for partial financial support (304706/2023-0).

\section*{Disclaimer}

The views expressed in this paper are those of the authors and should not be interpreted as representing the positions of the Banco Central do Brasil or its board members.




\clearpage

\appendix



		\section{Maximum likelihood estimator for binned data} \label{sec:app_mle}

		Following a notation similar to  \citet{virkar_clauset2014power}, let $B = \{b_1,b_2,...,b_m\}$ be a set of bin boundaries, $b_1=0$, $b_i > 0$ for $i \in \{2,...,m\}$, and $b_j > b_i$ for $j>i$ and $i,j \in \{1,2,...,m\}$. With these boundaries, we define $m$ bins, with $[b_i,b_{i+1})$ being the $i$-th bin, $i \in \{1,2,...,m-1\}$, and $[b_m,+\infty)$ being the $m$-th bin. Denote by $H \in \{h_1,h_2,...,h_m\}$ the set of bin counts, such that $h_i$ is the number of raw observations in the $i$-th bin, $i = 1,2,...,m$. Lastly, let $n \equiv \sum_{i=j}^{m} h_i$ be the number of firms with at least $b_j = \underline{s} > 0$ employees.
  
		Suppose $S$, $S \geq \underline{s} = b_j>0$, follows a certain distribution. Given that, the log-likelihood function for the binned data over this support is	
		\begin{align} \label{eq:logl}
			\mathcal{L} = & \ln \left[P(S\geq b_m)^{h_m} P(\underline{s} \leq S < b_{j+1})^{h_j} \prod_{i=j+1}^{m-1} P(b_i \leq S < b_{i+1})^{h_i}  \right] \notag \\
			\mathcal{L} = & h_m \ln P(S\geq b_m) + h_j \ln \left[1-P(S\geq b_{j+1})\right] + \sum_{i=j+1}^{m-1} h_i \ln P(b_i \leq S < b_{i+1})
		\end{align}
		which allows us to get the maximum likelihood estimator (MLE) of the distributional parameters. One possibility is to numerically maximize the log-likelihood function \eqref{eq:logl}. Nevertheless, a computationally faster way is to derive and numerically solve the associated First-Order Conditions (FOCs), derived for Pareto and lognormal distributions in the following.  
		
		\subsection{Pareto distribution} \label{sec:app_mle_pareto}
		
		If $S$, $S \geq \underline{s} = b_j>0$, is Pareto distributed with shape parameter $k>0$,
		\begin{align} 
			P(S\geq b_i) = & \left(\underline{s}/b_i\right)^k \text{ for } i = j,j+1,...,m \notag \\
			P(b_i \leq S < b_{i+1}) = & P(S\geq b_i)-P(S\geq b_{i+1}) = \underline{s}^k\left(b_i^{-k}-b_{i+1}^{-k}\right) \text{ for } i = j,j+1,...,m-1 \notag
		\end{align}	
		Plugging these probabilities into the log-likelihood function \eqref{eq:logl},
		\begin{align} \label{eq:logl_pareto}
			\mathcal{L} = & h_m k \ln \underline{s} - h_m k \ln b_m  + \sum_{i=j}^{m-1} \left[ h_i k \ln \underline{s} +h_i \ln \left(b_i^{-k}-b_{i+1}^{-k}\right) \right] \notag  \\
			\mathcal{L} = & n k \ln \underline{s} - h_m k \ln b_m  + \sum_{i=j}^{m-1} h_i \ln \left(b_i^{-k}-b_{i+1}^{-k}\right)  
		\end{align}
		From \eqref{eq:logl_pareto}, which is equivalent to equation (3.1) of  \citet{virkar_clauset2014power}, one can obtain the desired FOC:
		\begin{align} 
			\frac{\partial \mathcal{L}}{\partial k} = & n \ln \underline{s} - h_m \ln b_m  - \sum_{i=j}^{m-1} h_i  \left[\frac{b_i^{-k} \ln b_i- b_{i+1}^{-k}\ln b_{i+1}}{b_i^{-k}-b_{i+1}^{-k}}\right] = 0 \label{eq:foc_k}
		\end{align}
		
		\subsection{Lognormal distribution} \label{sec:app_mle_lognormal}
		
		If $S - \underline{s} = S - b_j$, $ S - b_j >0$, is lognormally distributed with parameters $\mu$ and $\sigma>0$,
		\begin{align}
			P(S\geq b_i) = & \frac{1-\erf\left(z_i\right)}{2} \text{ for } i = j+1,j+2,...,m \notag \\
			P(b_i \leq S < b_{i+1}) = & P(S\geq b_i)-P(S\geq b_{i+1}) = \frac{\erf\left(z_{i+1}\right)-\erf\left(z_i\right)}{2} \text{ for } i = j+1,...,m-1 \notag
		\end{align}	
		where $z_i \equiv \frac{\ln (b_i-\underline{s}) - \mu}{\sigma \sqrt{2}}$ and $\erf(z) \equiv \frac{2}{\sqrt{\pi}}\int_0^z e^{-t^2}dt$ is the Gaussian error function. Plugging these probabilities into the log-likelihood function \eqref{eq:logl},
		\begin{align} \label{eq:logl_lognormal}
			\mathcal{L} = & h_m\ln\left[\frac{1-\erf\left(z_m\right)}{2}\right] + h_j\ln\left[\frac{1+\erf\left(z_{j+1}\right)}{2}\right] + \sum_{i=j+1}^{m-1} h_i \ln \left[\frac{\erf\left(z_{i+1}\right)-\erf\left(z_i\right)}{2}\right] 
		\end{align}
		The FOCs for the maximization of the log-likelihood function \eqref{eq:logl_lognormal} are
		\begin{align} 
			\frac{\partial \mathcal{L}}{\partial \mu} = & h_m \frac{ \erf'(z_m) }{P(S\geq b_m)2\sigma\sqrt{2}} - h_j \frac{ \erf'(z_{j+1}) }{P(S<b_{j+1})2\sigma\sqrt{2}} + \sum_{i=j+1}^{m-1} h_i \frac{ \erf'(z_i)-\erf'(z_{i+1}) }{ P(b_i \leq S < b_{i+1})2\sigma\sqrt{2}} = 0 \notag \\
			& h_m \frac{ e^{-z_m^2} }{P(S\geq b_m)} - h_j \frac{ e^{-z_{j+1}^2} }{P(S<b_{j+1})} + \sum_{i=j+1}^{m-1} h_i \frac{ e^{-z_i^2}-e^{-z_{i+1}^2} }{ P(b_i \leq S < b_{i+1})} = 0 \label{eq:foc_mu} \\
			\notag \\
			\frac{\partial \mathcal{L}}{\partial \sigma} = & h_m \frac{ \erf'(z_m) z_m}{P(S\geq b_m)2\sigma} - h_j \frac{ \erf'(z_{j+1}) z_{j+1}}{P(S<b_{j+1})2\sigma} + \sum_{i=j+1}^{m-1} h_i \frac{ \erf'(z_i) z_i-\erf'(z_{i+1}) z_{i+1}}{P(b_i \leq S < b_{i+1})2\sigma} = 0 \notag \\
			& h_m \frac{ z_m e^{-z_m^2}}{P(S\geq b_m)} - h_j \frac{ z_{j+1} e^{-z_{j+1}^2}}{P(S<b_{j+1})} + \sum_{i=j+1}^{m-1} h_i \frac{ z_i e^{-z_i^2}-z_{i+1} e^{-z_{i+1}^2}}{P(b_i \leq S < b_{i+1})} = 0 \label{eq:foc_sigma}
		\end{align}
		where we use $\erf'(z) = \frac{2}{\sqrt{\pi}}e^{-z^2}$ to get each condition.


\clearpage

\bibliographystyle{agsm}

\bibliography{bibliografia}

\end{document}